\documentclass[conference,a4paper]{IEEEtran}
\IEEEoverridecommandlockouts
\usepackage{amsmath,amssymb,amsfonts}
\usepackage{algorithmic}
\usepackage{graphicx}
\usepackage{dblfloatfix}
\usepackage{textcomp}
\usepackage{xcolor}
\usepackage{float}
\usepackage{multirow}
\usepackage{footnote}

\usepackage{tikz}
\usepackage{pgfplots}
\pgfplotsset{compat=1.18}

\usepackage{csquotes}
\usepackage{setspace}
\usepackage{subcaption}
\usepackage{booktabs}

\usepackage{acronym}

\newacro{5g}[5G]{Fifth Generation}
\newacro{6g}[6G]{Sixth Generation}
\newacro{as}[AS]{Authentication Server}
\newacro{atm}[ATM]{Automated Teller Machine}
\newacro{ai}[AI]{Artificial Intelligence}
\newacroplural{ais}[AIs]{Artificial Intelligences}
\newacro{arima}[ARIMA]{Autoregressive Integrated Moving Average}
\newacro{AcCoRD}[AcCoRD]{Actor-based Communication via Reaction-Diffusion}
\newacro{b5g}[B5G]{Beyond 5G}
\newacro{ban}[BAN]{Body Area Network}
\newacro{bsi}[\textit{BSI}]{\textit{Federal Office for Information Security}}
\newacro{bdr}[BDR]{Bit Disagreement Rate}
\newacro{bs}[BS]{Base Station}
\newacro{ca}[CA]{Certification Authority}
\newacro{cav}[CAV]{Connected Autonomous Vehicles}
\newacro{cc}[CC]{Common Criteria}
\newacro{cir}[CIR]{Channel Impulse Response}
\newacro{cr}[CR]{Challenge-Response}
\newacro{cpu}[CPU]{Central Processing Unit}
\newacro{cpps}[CPPS]{Cyber-Physical Production System}
\newacro{crl}[CRL]{Certificate Revocation List}
\newacro{csi}[CSI]{Channel State Information}
\newacro{crke}[CRKE]{Channel-Reciprocity Based Key Extraction}
\newacro{ctf}[CTF]{Channel Transfer Function}
\newacro{cotf}[COTF]{Commercial-off-the-Shelf}
\newacro{cmos}[CMOS]{Complementary Metal-Oxide-Semiconductors}
\newacro{cfd}[CFD]{Computational Fluid Dynamics}
\newacro{dos}[DoS]{Denial-of-Service}
\newacro{ddos}[DDoS]{Distributed-Denial-of-Service}
\newacro{dna}[DNA]{Deoxyribonucleic Acid}
\newacro{dtls}[DTLS]{Datagram Transport Layer Security}
\newacro{dct}[DCT]{Discrete Cosine Transformation}
\newacro{dlt}[DLT]{Distributed Ledger Technology}
\newacro{dwt}[DWT]{Discrete Wavelet Transform}
\newacro{dmppic}[DMPPIC]{Diffusive Multi-Phase Particle-In-Cell} 
\newacro{eal}[EAL]{Evaluation Assurance Level}
\newacro{ecc}[ECC]{Elliptic Curve Cryptography}
\newacro{ecg}[ECG]{Electrocardiogram}
\newacro{eeg}[EEG]{Electroencephalogram}
\newacro{embb}[eMBB]{enhanced Mobile broad-Band}
\newacro{emg}[EMG]{Electromyogram}
\newacro{eog}[EOG]{Electrooculography}
\newacro{enb}[eNodeB]{Evolved Node B}
\newacro{er}[ER]{Extended Reality}
\newacro{fpga}[FPGA]{Field Programmable Gate Array}
\newacro{fdd}[FDD]{Frequency Division Duplexing}
\newacro{gdpr}[GDPR]{General Data Protection Regulation}
\newacro{gd}[G\&D]{Giesecke \& Devrient}
\newacro{h2m}[H2M]{Human-to-Machine}
\newacro{h2s}[H2S]{Human-to-Service}
\newacro{hmac}[HMAC]{Keyed-Hash Message Authentication Code}
\newacro{htc}[HTC]{Hologaphic-Type Communication}
\newacro{hotp}[HOTP]{HMAC-based One-time Password Algorithm}
\newacro{hsm}[HSM]{Hardware Security Module}
\newacro{ics}[ICS]{Industrial Control System}
\newacro{isi}[ISI]{Inter-Symbol Interference}
\newacro{iacs}[IACS]{Industrial Automation and Control System}
\newacro{ioe}[IoE]{Internet of Everything}
\newacro{iiot}[IIoT]{Industrial Internet of Things}
\newacro{iot}[IoT]{Internet of Things}
\newacro{iobnt}[IoBNT]{Internet of Bio-Nano Things}
\newacro{io}[I/O]{Input/Output}
\newacro{ic}[IC]{Integrated Circuit}
\newacro{id}[ID]{Identificator}
\newacro{ids}[IDS]{Intursion Detection System}
\newacro{irs}[IRS]{Intelligent Reflecting Surface}
\newacro{istn}[ISTN]{Integrated Space and Terrestrial Network}
\newacro{it}[IT]{Information Technology}
\newacro{itu}[ITU]{International Telecommunication Union}
\newacro{jcop}[JCOP]{Java Card Open Platform}
\newacro{kba}[KBA]{Knowledge Based Authentication}
\newacro{kdf}[KDF]{Key Derivation Function}
\newacro{kf}[KF]{Kalman Filter}
\newacro{led}[LED]{Light Emitting  Diode}
\newacro{lte}[LTE]{Long Term Evolution}
\newacro{ltea}[LTE-A]{Long Term Evolution Advanced}
\newacro{lr}[LR]{Linear Regression}
\newacro{los}[LoS]{Line of Sight}
\newacro{lorawan}[LoRaWAN]{Long Range Wide Area Network}
\newacro{mbb}[MBB]{Mobile Broadband}
\newacro{mc}[MC]{Molecular Communication}
\newacro{mfa}[MFA]{Multi-Factor Authentication}
\newacro{mcc}[MCC]{Mobile Cloud Computing}
\newacroplural{mcus}[MCUs]{Microcontroler Units}
\newacro{m2m}[M2M]{Machine-to-Machine}
\newacro{m2s}[M2S]{Machine-to-Service}
\newacro{mimo}[MIMO]{Multiple Input Multiple Output}
\newacro{mmimo}[mMIMO]{massive Multiple Input Multiple Output}
\newacro{ml}[ML]{Machine Learning}
\newacro{mulc}[mULC]{massive Ultra-Reliable Low-Latency Communication}
\newacro{mmtc}[MMTC]{massive Machine Type Communication}
\newacro{mmg}[MMG]{Mechanomyogram}
\newacro{multos}[MULTOS]{Multii-Application Smart Card Operating System}
\newacro{mux}[MUX]{Multiplexer}
\newacro{mnc}[MNC]{Mobile Network Code}
\newacro{me}[ME]{Mobile Environment}
\newacro{mac}[MACs]{Message Authentication Codes}
\newacro{mps}[MPS]{Master Production Schedule}
\newacro{mppic}[MPPIC]{Multi-Phase Particle-In-Cell} 
\newacro{ngmn}[NGMN]{Next Generation Mobile Network}
\newacro{nic}[NIC]{Network Interface Controller}
\newacro{nist}[NIST]{National Institute of Standards and Technology}
\newacro{oath}[OATH]{Open Authentication}
\newacro{ocra}[OCRA]{\ac{oath} Challenge-Response Algorithm}
\newacro{ocsp}[OCSP]{Online Certificate Status Protocol}
\newacro{otp}[OTP]{One-Time Password}
\newacro{ook}[OOK]{On Off Keying}
\newacro{pvc}[PVC]{Polyvinyl chloride}
\newacro{pa}[PA]{Process Automation}
\newacro{pap}[PAP]{Password-Authentication-Protocol}
\newacro{physec}[PhySec]{Physical Layer Security}
\newacro{pfs}[PFS]{Perfect Forward Secrecy}
\newacro{pin}[PIN]{Personal Identification Number}
\newacro{pkc}[PKC]{Public Key Cryptography}
\newacro{pki}[PKI]{Public Key Infrastructure}
\newacro{ppg}[PPG]{Photoplethysmography}
\newacro{prng}[PRNG]{Pseudo Random Number Generator}
\newacro{puf}[PUF]{Physically Unclonable Function}
\newacroplural{pufs}[PUFs]{Physically Unclonable Functions}
\newacro{pla}[PLA]{Physical Layer Authentication}
\newacro{plc}[PLC]{Programmable Logic Controller}
\newacro{qr}[QR]{Quick Response}
\newacro{rat}[RAT]{Radio Access Technology}
\newacro{radius}[RADIUS]{Remote Authentication Dial-In User Service}
\newacro{ram}[RAM]{Random-Access Memory}
\newacro{ran}[RAN]{Radio Access Networks}
\newacro{rf}[RF]{Radio-Frequency}
\newacro{rfid}[RFID]{Radio-Frequency Identification}
\newacro{ris}[RIS]{Reconfigurable Intelligent Surface}
\newacro{rng}[RNG]{Random Number Generator}
\newacro{ro}[RO]{Ring-Oscillator}
\newacro{rom}[ROM]{Read-Only Memory}
\newacro{rs}[RS]{Reed-Solomon}
\newacro{rsa}[RSA]{Rivest-Shamir-Adleman}
\newacro{rssi}[RSSI]{Received Signal Strength Indicator}
\newacro{rsrp}[RSRP]{Reference Signal Received Power}
\newacro{re}[RE]{Resource Elements}

\newacro{maf}[MAF]{Moving Average Filter}
\newacro{snr}[SNR]{Signal-to-Noise Ratio}
\newacro{sdn}[SDN]{Software-Defined Network}
\newacro{sdr}[SDR]{Software-Defined Radio}
\newacro{seccos}[SECCOS]{Secure Chip Card Operating System}
\newacro{sip}[SIP]{Session Initiation Protocol}
\newacro{skg}[SKG]{Secret Key Generation}
\newacro{sram}[SRAM]{Static Random Access Memory}
\newacro{srs}[SRS]{Software Radio Systems}
\newacro{starcos}[STARCOS]{Smart Card Chip Operating System}
\newacro{sha}[SHA]{Secure Hash Algorithm}
\newacro{se}[SE]{Static Environment}
\newacro{svm}[SVM]{Support Vector Machine}
\newacro{tcg}[TCG]{Trusted Computing Group}
\newacro{tpm}[TPM]{Trusted Platform Module}
\newacro{tls}[TLS]{Transport Layer Security}
\newacro{trng}[TRNG]{True Random Number Generator}
\newacro{tsn}[TSN]{Time-Sensitve Networking}
\newacro{tofu}[TOFU]{Trust On First Use}
\newacro{tufu}[TUFU]{Trust Upon First Use}
\newacro{totp}[TOTP]{Time-based One-time Password Algorithm}
\newacro{tia}[TIA]{Totally Integrated Automation}
\newacro{uav}[UAV]{Unmanned Arial Vehicles}
\newacro{usb}[USB]{Universal Serial Bus}
\newacro{usrp}[USRP]{Universal Software Radio Peripheral}
\newacro{uhd}[UHD]{USRP Hardware Driver}
\newacro{usim}[USIM]{Universal Subscriber Identity Module}
\newacro{ue}[UE]{User Equipment}
\newacro{urllc}[URLLC]{Ultra-Reliable Low-Latency Communication}
\newacro{ulbc}[ULBC]{Ultra-Reliable Low-Latency Broadband Communication}
\newacro{umbb}[uMBB]{ubiquious Mobile Broadband}
\newacro{ummimo}[UM-MIMO]{Ultra-Massive MIMO}
\newacro{vlc}[VLC]{Visible Light Communication}
\newacro{warp}[WARP]{Wireless open-Access Research Platform}
\newacro{wt}[WT]{Wavelet Transform}

\usepackage[
	pdfborder={0 0 0},
	colorlinks=false,% 
	urlcolor=black,%
	linkcolor=black,%
	citecolor=black,%
	filecolor=black,%
	breaklinks,%
	]{hyperref}%
	
\usepackage{cite} 
\usepackage{cite}  % Optional, for better citation formatting in IEEE style

\def\BibTeX{{\rm B\kern-.05em{\sc i\kern-.025em b}\kern-.08em T\kern-.1667em\lower.7ex\hbox{E}\kern-.125emX}}
    
\begin{document} 

\title{
A Numerical and Experimental Evaluation of Microbubble Communication Using OpenFOAM
\thanks{
%This work has been supported by the Federal Ministry of Research, Technology and Space of the Federal Republic of Germany (F\"{o}rderkennzeichen 16KIS1990, IoBNT) in Cooperation with (F\"{o}rderkennzeichen 16KIS2402K, Open6GHub+). The authors alone are responsible for the content of the paper. 
This research was supported by the German Federal Ministry of Research, Technology and Space (BMFTR) within the projects IoBNT and Open6GHub+ under grant numbers 16KIS1990 and 16KIS2402K. The responsibility for this publication lies with the authors.
This is a preprint of a work accepted but not yet published at the 30th ITG-Symposium, Mobile Communications - Technologies and Applications in Osnabrueck, Germany. }
} 

\author{
\IEEEauthorblockN{Annika Tjabben\IEEEauthorrefmark{1}, Carolin Conrad\IEEEauthorrefmark{1}, and Hans D. Schotten\IEEEauthorrefmark{1}\IEEEauthorrefmark{2}} 
\IEEEauthorblockA{\IEEEauthorrefmark{1}German Research Center for Artificial Intelligence (DFKI), Kaiserslautern, Germany} 
\IEEEauthorblockA{\IEEEauthorrefmark{2}Department of Electrical and Computer Engineering, RPTU University Kaiserslautern-Landau, Kaiserslautern, Germany\\ E-mail: \{Annika.Tjabben, Carolin.Conrad, Hans.Schotten\}@dfki.de} 

}

\maketitle
%\\ E-mail: Schotten@rptu.de
% 
%++++++++++++++++++++++++++++++++++++++++++++++++++++++++++++++++++++++++++++++++++++++++++++++++++++
%
%
%		ABSTRACT
%
%Reliable communication in confined environments -- like blood vessels or industrial pipelines -- remain challenging due to signal attenuation and limited sensor accessibility. Therefore, this work investigates microbubbles as robust information carriers within the Internet of Bio-Nano Things (IoBNT) paradigm, leveraging their established use as ultrasound contrast agents. \tbd{Bezug zu 6G erwähnen}

%$This work presents a combined experimental and numerical analysis characterizing microbubble transport under varying flow conditions relevant to biomedical and industrial applications. Experiments with \textit{SonoVue} microbubbles in a recirculating water channel validate an OpenFOAM-based Computational Fluid Dynamics (CFD) simulation using the \textit{incompressibleDenseParticleFluid} solver. Key cases examine water vs. blood-like media and high vs. physiological flow velocities, analyzing the relative influence of fluid properties and advection on microbubble dynamics. Recirculation effects are considered in relation to in vivo circulation timescales.
%
%++++++++++++++++++++++++++++++++++++++++++++++++++++++++++++++++++++++++++++++++++++++++++++++++++++
\begin{abstract}
Reliable communication in confined environments, such as blood vessels or industrial pipelines, remain challenging due to signal attenuation and limited sensor accessibility. 
% In this context, future 6G/6G+ systems are expected to support integrated sensing and novel bio-digital interfaces, motivating communication paradigms for the Internet of Bio-Nano Things (IoBNT). 
Therefore, this work investigates microbubbles as robust information carriers within the Internet of Bio-Nano Things (IoBNT) paradigm, leveraging their established use as ultrasound contrast agents. 
It presents a combined experimental and numerical analysis characterizing microbubble transport under varying flow conditions relevant to biomedical and industrial applications. Experiments with SonoVue microbubbles in a recirculating water channel validate an OpenFOAM-based Computational Fluid Dynamics (CFD) simulation using the incompressibleDenseParticleFluid solver. Key cases examine water vs. blood-like media and high vs. physiological flow velocities, analyzing the relative influence of fluid properties and advection on microbubble dynamics. Recirculation effects are considered in relation to in vivo circulation timescales. 
\end{abstract}

\section{Introduction} 
\label{sec:introduction}

\noindent Reliable communication within confined or inaccessible environments remains a fundamental challenge in both biomedical and industrial contexts. In healthcare applications, conventional electromagnetic communication techniques, particularly within the terahertz frequency range, are difficult to apply inside the human body due to strong attenuation and limited penetration depth in biological tissues. Consequently, alternative communication paradigms are required to enable reliable data exchange under such constraints, as envisioned in the emerging field of the \ac{iobnt}~\cite{Akyildiz2015,lee2023internet}. In this context, \ac{iobnt} conceptual developments and technologies are considered part of next-generation 6G networks and beyond~\cite{Akyildiz2020}. 

A promising approach in this context is the use of microbubbles as signaling molecules. Microbubbles are gas-filled particles containing sulfur hexafluoride ($\text{SF}_6$), encapsulated by a phospholipid shell. They are well established in clinical practice as ultrasound contrast agents. There they enhance image quality during diagnostic examinations. Beyond imaging, microbubbles have also been investigated for disease detection and targeted drug or gene delivery, demonstrating their versatility, controllability and biocompatibility. 

The acoustic properties of microbubbles make them attractive candidates for information transfer within biological media. They can be excited and detected using ultrasound waves within specific frequency ranges. Low-frequency ultrasound can be employed to monitor their motion, whereas high-frequency excitation induces cavitation and bubble collapse, releasing encapsulated substances. This mechanism has been exploited for localized drug or gene release, such as in targeted chemotherapy, minimizing side effects by activating delivery only in affected tissues. Furthermore, recent work has shown that microbubbles can be acoustically steered against fluid flow, enabling controlled propagation even in vascular environments~\cite{DelCampoFonseca2023}.

Building on a previously developed graph-based simulation framework for modeling microbubble propagation in vascular networks, this work advances toward a physics-based OpenFOAM implementation that captures fluid–particle interactions and transport dynamics under realistic flow conditions~\cite{Tjabben2024}. This extension enables investigation of microbubble-based communication performance in complex flow environments, bridging theoretical modeling and experimental validation. 

In addition to biomedical applications, microbubble-based communication can also be extended to industrial scenarios. In liquid-filled or metallic environments, such as pipes in chemical processing plants or cooling systems, traditional electromagnetic communication suffers from severe attenuation, while wired sensors require costly installation and maintenance. In such cases, microbubbles introduced into the fluid at specific intervals (e.g., in pulse patterns suitable for \ac{ook} modulation) can serve as discrete carriers of information. Ultrasonic sensors positioned along the pipeline can detect their presence, frequency, or timing to extract transmitted information. Moreover, analyzing missing or delayed microbubble signals could enable fault detection, such as identifying leaks or flow irregularities.

The objective of using microbubbles as information carriers is not to transmit large data volumes but to enable low-rate, robust, and secure signaling in environments where conventional communication channels fail. Such methods could support the transmission of critical control messages or sensitive data, including encryption keys, within otherwise inaccessible areas.

The remainder of this work is organized as follows: \autoref{sec:state} provides an overview of related work in molecular and acoustic communication. \autoref{sec:setup} introduces the experimental setup and describes the simulation framework developed based on the measurement environment. \autoref{sec:measurments} discusses the conducted measurements and relevant parameter selection. \autoref{sec:results} presents the results along with a comprehensive discussion. 
Finally, \autoref{sec:conlusion} concludes the paper and outlines future research directions.

\section{Related Work} 
\label{sec:state}
\noindent Molecular communication has emerged as a promising paradigm for enabling information exchange in environments where conventional electromagnetic communication is infeasible, which has lead to extensive research on system models, simulation frameworks, and experimental implementations.

\textit{Farsad et al.} provide a comprehensive survey of molecular communication systems, including candidate biological, chemical, and physical processes for transmitters, receivers, and propagation mechanisms, and reviewing key system models, modulation and detection techniques, experimental implementations, and open research challenges for future molecular communication networks~\cite{Farsad2016}.

To support performance evaluation of diffusion-based links, \textit{Yilmaz and Chae} developed a programmable end-to-end simulator for molecular communication via diffusion in \mbox{one-,} \mbox{two-,} and three-dimensional environments~\cite{YILMAZ2014136}. Their framework enables the generation of sequential symbol transmissions and incorporates physical channel properties to analyze \ac{isi}, showing that the simulation runtime scales linearly with the number of transmitted signals and that \ac{isi} can be mitigated by decision-feedback detection~\cite{YILMAZ2014136}. Building on this work, \textit{Yilmaz et al.} also derived a closed-form analytical model for a three-dimensional diffusion channel with a fully absorbing spherical receiver, characterizing the time-varying hitting probability and the cumulative number of absorbed molecules as functions of system parameters~\cite{yilmaz2014}. These models provide a foundation for understanding signal attenuation and delay in diffusion-dominated molecular channels.

Beyond purely analytical and custom-built simulators, general-purpose \ac{cfd} tools have recently been employed to study particle-based molecular communication in more complex geometries. \textit{Hofmann et al.} employed the \ac{mppic} solver in OpenFOAM to model molecular communication with discrete particles transported in laminar flow, and validated the numerical results against an analytical reference and a MATLAB-based particle simulation~\cite{hofmann2024openfoam}. The resulting channel impulse responses were in good agreement, with only minor deviations at larger transmitter–receiver distances. Extending this approach, \textit{Zhou et al.} introduced a \ac{dmppic} solver that augments \ac{mppic} with random walk dynamics to incorporate Brownian motion~\cite{zhou2024diffusive}. 

In parallel, several specialized reaction-diffusion simulators have been developed with molecular communication applications in mind. The \ac{AcCoRD} framework is a hybrid microscopic/mesoscopic simulator implemented in C with MATLAB utilities~\cite{Noel2017}. It allows flexible three-dimensional environment configurations, includes reactions, flow, and boundary interactions, and is designed to efficiently generate channel statistics such as receiver observation distributions. In contrast, biochemical simulators such as Smoldyn provide highly detailed stochastic particle-based reaction-diffusion modeling for cell biology, but are not primarily tailored to communication-system abstraction or metrics~\cite{Andrews2010Smoldyn}. These tools are therefore more suitable when intracellular reaction networks are of interest rather than end-to-end communication performance evaluation. 

\section{Testbed} % and Measurements 
\label{sec:setup}

\begin{figure} [h]
    \centering
    \includegraphics[width=0.95\linewidth]{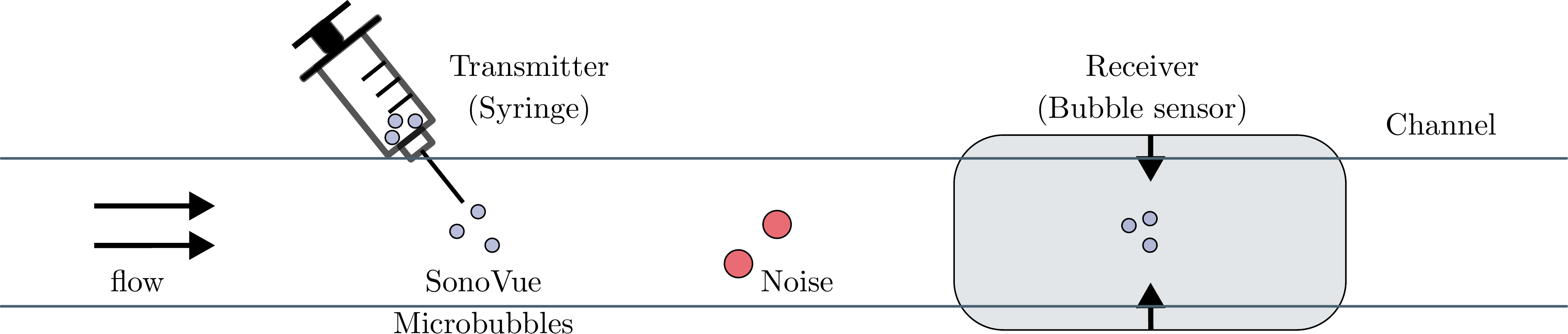}
    \caption{Schematic representation of data transmission via microbubbles~\cite{Tjabben2026}.}
    \label{img:channel}
\end{figure}

\begin{figure} [h]
    \centering
    \includegraphics[width=0.6\linewidth]{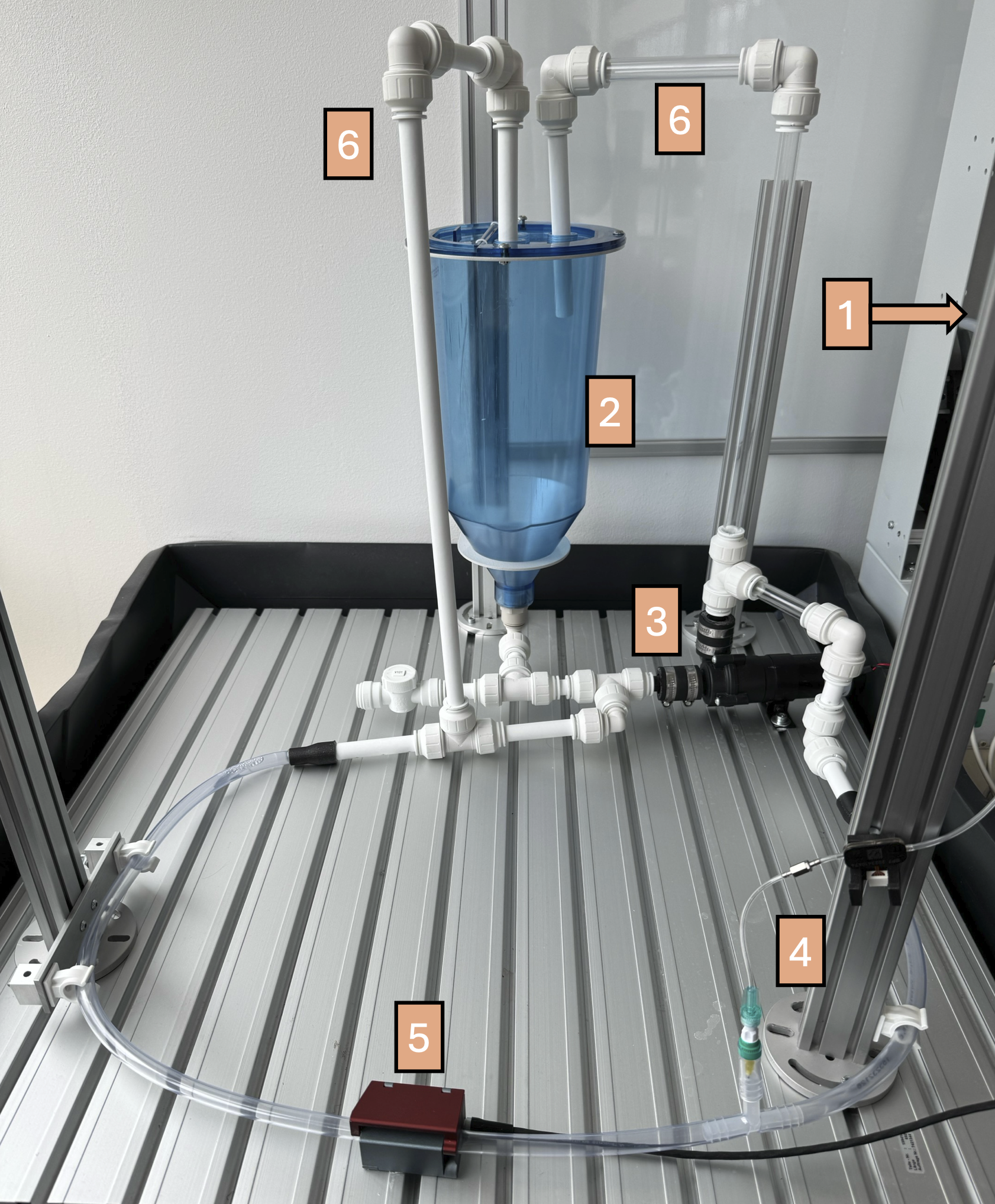} 
    \caption{Experimental setup for measurements. Components: (1) Siemens S7-1500 PLC, (2) water tank, (3) water pump, (4) micropump, (5) bubble sensor, (6) aerating pipes)}
    \label{img:exp_setup}
\end{figure}

\noindent In this section, the experimental setup is described. The setup is based on a transmitter-receiver system. A syringe serves as the transmitter, injecting microbubbles as information carriers into the system, while an ultrasound sensor acts as the receiver. A schematic overview of this system is shown in Fig.~\ref{img:channel}. The experimental setup and its description are based on the work of \textit{Tjabben et al.}~\cite{Tjabben2026} and are shown in Fig.~\ref{img:exp_setup}. In the following, the main components retained from the original setup are briefly summarized, and the newly introduced elements are described. 

The system structure remains based on 3/8\,inch tubing, arranged as a continuous loop to enable microbubble recirculation. The benefits of enhanced visualization and improved representation of biological flow conditions are thus retained. Distilled water is used as the working medium. A stable flow is generated by the SPX Flow CM30P7-1 pump, operated at 3\,V during measurements. This results in a reduced flow rate of 1,51\,L/min lower than typical rates in the human blood circulation. The reduced flow rate is necessary due to the larger tubing dimensions, as the ultrasonic sensor is compatible only with this scale. System operation and monitoring are managed by a Festo Master Production Schedule Process Automation Compact-Workstation in conjunction with a Siemens S7-1500 Programmable Logic Controller monitor, as shown in Fig.~\ref{img:exp_setup}. 
As mentioned above, microbubbles are used as signal molecules, more precisely SonoVue microbubbles from BRACCO. Typically used as a medical contrast agent for ultrasound imaging, these microbubbles are slightly soluble in water and effectively reflect sound waves at the water interface. Each millilitre of the solution contains approximately 8\,\textnormal{\textmu{}L} of microbubbles, each with an average diameter of 2.5\,\textnormal{\textmu{}m} per microbubble~\cite{sonovue2021}. For the measurements, the solution is diluted with a saline solution to adjust the concentration of microbubbles per injection in the ratio of 0.1\,mL microbubble solution to 2\,mL of saline solution. The injection is performed with an Arduino-controlled motorized syringe to achieve a uniform and precise injection, generating a ``high'' signal. In combination with the GAMPT BubbleCounter BCF300, a Doppler ultrasound sensor used as the receiver, the microbubbles are clearly detectable. The sensor, designed for air bubble analysis in extracorporeal circulation systems, is mounted on a PVC tube downstream of the injection point.\\ 
Several improvements to the experimental setup were implemented. A major issue was the presence of air bubbles affecting the measurements. To remove them more efficiently, additional tubing was added near the pump and downstream between the measurement point and the pump. The injection method was also redesigned. Instead of the Arduino-controlled motorized syringe, a Bartels BP7-Tubing Micropump with an mp-Highdriver Pump Driver is now used to inject the diluted microbubble solution enabling a more stable and precise injection process. 

The simulation of the experimental setup was developed using OpenFOAM, an open-source framework for \ac{cfd} that enables three-dimensional flow simulations. OpenFOAM allows the visualization and analysis of complex flow phenomena such as turbulence, chemical reactions, and thermal conduction~\cite{openfoam2026}. The numerical results were visualized, animated, and further analyzed using the open-source software Paraview. 

\begin{figure} [t!]
    \centering
    \includegraphics[width=0.95\linewidth]{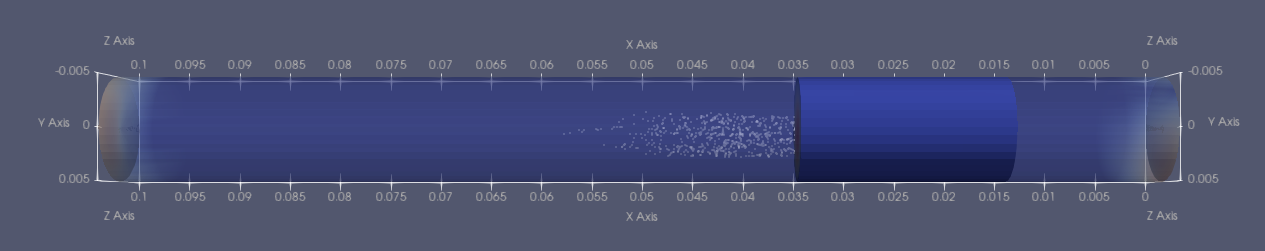} 
    \caption{Final Paraview simulation showing microbubbles within a transparent tube. Highlighted region indicates the measurement range.} 
    \label{img:simulation_receiver}
\end{figure}

The simulation parameters were defined based on the experimental setup, including the choice of solver and time control. Since the simulation involves distilled water and blood as media and requires modeling microbubbles as discrete particles, the solver ``incompressibleDenseParticleFluid'' was applied. The time control was adjusted to match the duration of a single measurement period. 
The mesh represents only the most relevant region of the experimental system, specifically, the communication section. 
This includes the microbubble injection area, the propagation medium, and the sensor region.
Microbubbles were modeled as particles corresponding to the average size of the SonoVue microbubbles, with a defined injection box representing the nozzle entry point. This box served as the source region for particles at the start of each simulated injection.
The medium was characterized by its density, viscosity model, and kinematic viscosity, while its flow conditions were defined by the experimentally measured flow rate and selected turbulence model. More detailed information about the physical properties of water and blood used in the simulation can be found in \autoref{sec:measurments}. 
The measurement range corresponding to the ultrasonic sensor was added during post-processing in Paraview and is shown in Fig.~\ref{img:simulation_receiver}. 
For this purpose, a 2\,cm wide area was defined 5\,cm downstream of the injection point, corresponding to the sensor’s position in the experimental setup. The dimensions were measured on the experimental setup. When microbubble particles enter this region during the simulation, they are counted and stored in a variable for subsequent analysis.

\section{Measurements} 
\label{sec:measurments}
\noindent This section describes the measurements performed to characterize the experimental setup and to validate the simulation model, as well as additional numerical studies not feasible experimentally.

\begin{figure}[h]
\centering
\begin{tikzpicture}
\begin{axis}[
    width=1\linewidth,
    height=5.2cm,
    xlabel={Time step},
    ylabel={Volume},
    xmin=0, xmax=200,
    ymin=0, ymax=4000,
    grid=major,
    enlargelimits=0.02
]

\addplot[
    thick,
    blue
] coordinates {
(0,15.9513) (1,42.9759) (2,29.4402) (3,12.7738) (4,158.43)
(5,141.32) (6,33.4929) (7,23.9843) (8,181.432) (9,112.48)
(10,51.7655) (11,79.6267) (12,270.388) (13,115.061) (14,75.6971)
(15,91.5801) (16,200.686) (17,140.497) (18,124.088) (19,115.588)
(20,361.097) (21,194.333) (22,136.697) (23,108.245) (24,528.267)
(25,298.083) (26,170.795) (27,172.277) (28,734.205) (29,373.85)
(30,235.835) (31,211.958) (32,738.718) (33,356.261) (34,250.168)
(35,252.399) (36,869.575) (37,409.056) (38,309.253) (39,296.362)
(40,611.657) (41,326.042) (42,314.415) (43,305.726) (44,1002.75)
(45,448.047) (46,347.493) (47,321.571) (48,932.921) (49,484.185)
(50,406.574) (51,540.21) (52,961.674) (53,415.896) (54,429.547)
(55,496.004) (56,1055.36) (57,477.804) (58,468.417) (59,703.938)
(60,1242.61) (61,590.128) (62,522.397) (63,768.339) (64,1021.71)
(65,603.668) (66,590.559) (67,770.591) (68,1069.88) (69,591.676)
(70,579.127) (71,677.633) (72,1122.3) (73,606.569) (74,600.116)
(75,844.12) (76,1149.47) (77,572.834) (78,709.709) (79,697.039)
(80,1542.87) (81,705.789) (82,682.943) (83,760.986) (84,1382.79)
(85,749.423) (86,680.662) (87,664.699) (88,1455.91) (89,752.362)
(90,820.198) (91,821.176) (92,1513.53) (93,904.21) (94,795.919)
(95,688.524) (96,2087.67) (97,875.816) (98,824.002) (99,791.19)
(100,1931.57) (101,901.938) (102,1055.55) (103,937.623)
(104,1851.29) (105,1014.66) (106,840.052) (107,794.645)
(108,1783.23) (109,959.807) (110,1096.06) (111,1009.01)
(112,2161.25) (113,1141.34) (114,952.116) (115,1088.22)
(116,1783.63) (117,1173.34) (118,1153.78) (119,1056.38)
(120,2006.65) (121,1224.94) (122,996.224) (123,1054.51)
(124,2136.26) (125,1089.57) (126,1151.82) (127,1059.91)
(128,1794.1) (129,1282.44) (130,1261.72) (131,1098.62)
(132,1770.1) (133,1179.68) (134,1264.09) (135,1281.02)
(136,1777.99) (137,1440.12) (138,1205.45) (139,1706.88)
(140,1986.44) (141,1452.3) (142,1187.85) (143,1639.21)
(144,1741.83) (145,1357.89) (146,1436.19) (147,1959.26)
(148,2104.3) (149,1295.39) (150,1320.96) (151,1882.15)
(152,1856.21) (153,1302.59) (154,1463.68) (155,2134.34)
(156,1949.11) (157,1468.57) (158,1466.18) (159,2116.35)
(160,2148.23) (161,1670.23) (162,1444.53) (163,2407.36)
(164,2209.03) (165,1641.95) (166,1817.96) (167,2744.38)
(168,2370.44) (169,1958.1) (170,2008.22) (171,2842.08)
(172,2750.86) (173,2255.91) (174,2692.22) (175,3883.3)
(176,3115.04) (177,2120.89) (178,2318.83) (179,3238.67)
(180,3212.27) (181,2992.72) (182,2379.87) (183,3726.82)
(184,3373.04) (185,2669.46) (186,2541.57) (187,3444.27)
(188,2943.34) (189,2669.53) (190,2394.1) (191,2981.65)
(192,2765.26) (193,2242.77) (194,2319.88) (195,3181.29)
(196,2506.35) (197,2306.18) (198,2208.15) (199,2767.03)
(200,2619.45)
};

\end{axis}
\end{tikzpicture}
\caption{Ultrasound data from Experimental 1, where the peaks indicate the presence of microbubbles.}
\label{fig:Exp1_Volumen}
\end{figure}
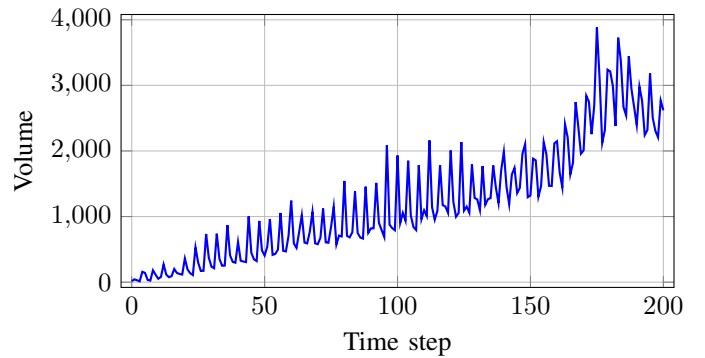

In experimental 1 (denoted as exp. 1 in \autoref{tab:blutgefaesse2}), a mixture of 0.1\,mL of SonoVue microbubble solution and 2\,mL of NaCl solution was injected into the flow channel; the corresponding ultrasound data are shown in Fig.~\ref{fig:Exp1_Volumen}.

The injections were performed periodically with an injection duration of 10\,ms followed by a pause of 1\,s, at a total volume flow of 1.5\,L/min. 
This corresponds to a mean flow velocity of approximately 0.351\,m/s in the pipe. Using a pipe inner diameter of 0.0095\,m and the kinematic viscosity of water at $20\,^\circ$C $\nu=1.002 \cdot 10^{-6}\,m^2/s$, 
the Reynolds number is $$Re = \frac{0.351\,\text{m}/\text{s} \cdot 0.0095\,\text{m}}{1.002\cdot 10^{-6}\,\text{m}^2/\text{s}} \approx 3327.844,$$ indicating a turbulent flow regime. 

Experimental 1 serves as the reference case for all subsequent investigations. To verify the simulation model, \mbox{numerical case 1} (num. 1) was conducted under identical conditions (i.e., water as the medium, a flow velocity of 0.351\,m/s, and a volume flow of 1.5\,L/min). 
In numerical case 2 (num. 2), the fluid properties were adjusted to represent blood, with a density of  $\rho = 1060\,\text{kg}/\text{m}^3$ and a kinematic viscosity of $\nu = 3.302\cdot 10^{-6}\,\text{m}^2/\text{s}$. The resulting Reynolds number is $$Re = \frac{0.351\,\text{m}/\text{s} \cdot 0.01\, \text{m}}{3.302 \cdot 10^{-6}\,\text{m}^2/\text{s}} \approx 1062.992,$$ which corresponds to laminar flow. 

The third measurement examines the influence of flow velocity. Here, flow conditions were adapted to approximate the velocities found in human arteries. Direct transfer of physiological velocities to the experimental setup is not possible, since typical vessel diameters are considerably smaller than the inner diameter of the pipe. This alters the volume flow and thus the resulting velocities required for a realistic representation. An artery is assumed to have an inner diameter of 4\,mm~\cite{blut2}. 
This results in a physiological volume flow of $$Q_{\text{real}} = v_{\text{blood}} A_{\text{blood}} = v_{\text{blood}} \frac{\pi}{4} d^2.$$ 
In the simulation, the pipe diameter remains at $d = 0.01\,\text{m}$ and $\nu = 1 \cdot 10^{-6}\,\text{m}^2/\text{s}$ is used as the kinematic viscosity for water. The Reynolds number for the flow conditions chosen to emulate an artery is $Re \approx 788.42$, which is in the laminar range. 
\autoref{tab:blutgefaesse2} provides an overview of all experimental and simulations conducted in this work. 

\begin{table}[h!]
    \centering
    \caption{Overview of all measurements with highlighted changed parameters.}
    \small
    \begin{tabular}{cccc}
    \toprule
    Measurement & Medium & Flow velocity & Volume flow  \\
     &  & [mm/s] & [L/min] \\
    \hline
    exp. 1 & water & $351$ & $1.5$ \\
    num. 1 & water & $351$ & $1.5$ \\
    num. 2 & blood & $351$ & $1.5$ \\
    num. 3 & water & $79$ & $0.3393$ \\
    %num. 4 & water & $\textbf{0,276}$ & $0,0012$ \\
    exp. 2 & water & $79$ & $0.3393$ \\
    \bottomrule
    \end{tabular}
    \label{tab:blutgefaesse2}
\end{table}

Due to the deterministic formulation of the underlying model, the numerical results exhibit a high degree of reproducibility, with deviations primarily attributable to discretization and solver configurations. In contrast, the experimental measurements are influenced by noise and environmental fluctuations, leading to minor deviations between individual experiments. Nevertheless, repeated experiments consistently show similar qualitative behavior, indicating the robustness of the observed communication properties.

\section{Results and Discussion} 
\input{./chapter/results.tex}

\section{Conclusion and Outlook}
\label{sec:conlusion}

\noindent This work presents an OpenFOAM-based simulation framework to assess the feasibility of microbubbles as information carriers for reliable communication in confined environments such as blood vessels and industrial pipelines. Experimental measurements using SonoVue microbubbles in a recirculating flow channel, combined with OpenFOAM simulations using the incompressibleDenseParticleFluid solver is used. This provides a characterization of transport dynamics under both laminar and turbulent conditions.

The results demonstrate that microbubbles can serve as viable signaling agents within the \acl{iobnt} framework, offering an energy-efficient alternative to electromagnetic communication in highly attenuating media. The proposed model provides a scalable foundation for designing secure, low-rate communication links in biomedical and industrial systems.

Future work will include implementing a fully closed-loop circulatory setup to study the behavior of recirculating microbubbles under realistic flow conditions. The integration of diffusion-based particle solvers will enhance the physical accuracy of nanoscale simulations. Finally, key communication metrics such as the bit error rate, channel capacity, and resilience against eavesdropping will be analyzed to quantify end-to-end system performance. 

%\clearpage
%++++++++++++++++++++++++++++++++++++++++++++++++++++++++++++++++++++++++++++++++++++++++++++++++++++
%
%
%		ACKNOWLEDGMENT
%
%
%++++++++++++++++++++++++++++++++++++++++++++++++++++++++++++++++++++++++++++++++++++++++++++++++++++   
%\section*{Acknowledgment}
%\input{./chapter/acknowledgement.tex}

%++++++++++++++++++++++++++++++++++++++++++++++++++++++++++++++++++++++++++++++++++++++++++++++++++++
%
%
%		REFERENCES
%
%
%++++++++++++++++++++++++++++++++++++++++++++++++++++++++++++++++++++++++++++++++++++++++++++++++++++
%\enlargethispage{3\baselineskip}
%\nocite{*}

%\printbibliography 
%\bibliography{references}
%\renewcommand*{\bibfont}{\footnotesize}
%\bibliographystyle{IEEEtran}  % Choose the style, e.g., IEEEtran for IEEE papers
%\enlargethispage{3\baselineskip}
\bibliography{references}  % Your .bib file name (without the .bib extension)

\end{document}